\documentclass[preprint,showpacs,preprintnumbers,amsmath,amssymb,groupedaddress]{revtex4}

\def\GeVcSq {{\rm (GeV/}c)^2}

\usepackage{graphicx}
\usepackage{dcolumn}
\usepackage{bm}

\begin{document}

\preprint{}

\title{Double Spin Asymmetries $A_{NN}$ and $A_{SS}$ at $\sqrt{s}=200$~GeV in Polarized Proton-Proton Elastic Scattering at RHIC}
\author{ S.~B\"{u}ltmann}      
\author{I.~H.~Chiang}
\author{R.~E.~Chrien}
\author{A.~Drees}
\author{R.~L.~Gill}
\author{W.~Guryn}
\author{J.~Landgraf}
\author{T.~A.~Ljubi\v{c}i\v{c}}
\author{D.~Lynn}
\author{C.~Pearson}
\author{P.~Pile}
\author{A.~Rusek}
\author{M.~Sakitt}
\author{S.~Tepikian}
\author{K.~Yip}
\affiliation{Brookhaven National Laboratory, Upton, NY 11973, USA}
\author{J.~Chwastowski} 
\author{B.~Pawlik}
\affiliation{Institute of Nuclear Physics, Cracow, Poland}
\author{M. Haguenauer}
\affiliation{Ecole Polytechnique, 91128 Palaiseau Cedex, France}
\author{A.~A.~Bogdanov}
\author{S.~B.~Nurushev}
\author{M.~F.~Runtzo}
\author{M.~N.~Strikhanov}
\affiliation{Moscow Engineering Physics Institute, Moscow, Russia}
\author{I.~G.~Alekseev}
\author{V.~P.~Kanavets}
\author{L.~I.~Koroleva} 
\author{B.~V.~Morozov} 
\author{D.~N.~Svirida}
\affiliation{Institute for Theoretical and Experimental Physics, Moscow, Russia}
\author{A.~Khodinov}
\author{M.~Rijssenbeek} 
\author{L.~Whitehead}
\author{S.~Yeung}
\affiliation{Stony Brook University, Stony Brook, NY 11794, USA}
\author{K.~De}
\author{N.~Guler}
\author{J.~Li}
\author{N.~\"{O}zt\"{u}rk}
\affiliation{University of Texas at Arlington, Arlington, TX 76019, USA}
\author{A.~Sandacz}
\affiliation{So\l tan Institute for Nuclear Studies, Warsaw, Poland}
\affiliation{}

\collaboration{PP2PP}
\homepage{}

\begin{abstract}
We present the first measurements of the double spin asymmetries $A_{NN}$ and 
$A_{SS}$ at $\sqrt{s}=200$ GeV, obtained by the pp2pp experiment using polarized proton beams at the Relativistic Heavy Ion Collider (RHIC).  
The data were collected in the four momentum transfer $t$ range $0.01 \leq |t| \leq 0.03$ $\GeVcSq$. 
The measured asymmetries, which are consistent with zero, allow us to estimate
upper limits on the double helicity-flip amplitudes $\phi _2$ and $\phi _4$ at small $|t|$ as well as on the difference $\Delta \sigma _T$ between 
the total cross sections for transversely polarized protons with 
antiparallel or parallel spin orientations.

\end{abstract}
\maketitle
\section{\label{sec:intro}Introduction\protect\\}

In this paper we present the first measurements of the 
double spin asymmetries $A_{NN}$ and $A_{SS}$ 
in elastic scattering of polarized protons at RHIC at 
$\sqrt{s} = 200 \:\rm{GeV}$ and low $t$-range $0.01 \leq |t| \leq 0.03$ $\GeVcSq$. 
The experiment was carried out by the pp2pp Collaboration 
at the RHIC-Spin complex as a part
of the pp2pp experimental program \cite{guryn,PP2PPplb04,lynn,PP2PPplb06}. 
The present results complement our previously published results on the
analysing power $A_N$ \cite{PP2PPplb06}.
The center of mass energy exceeds by a 
factor of 10 the energies reached in previous measurements
of the elastic scattering spin parameters \cite{e704,e950,bravar,jetcal,carboncal}.

The kinematic region of our experiment is suitable for an
investigation of the spin dependence of diffractive scattering ---
the dominant mechanism at high energies. In particular, the physical motivation
of the present experiment is connected with two long-standing questions.
First, is $s$-channel helicity conserved in elastic scattering at
asymptotic energies, which is directly related to the question if
the Pomeron exchange contributes to the spin
effects (ref.~\cite{buttimore} and references therein)? Second, does the
hypothetical Odderon exist \cite{leader}?

The double spin asymmetry $A_{NN}$ is defined as the cross section asymmetry,
\begin{equation}
A_{NN} = \frac{\sigma ^{\uparrow \uparrow + \downarrow \downarrow} -
               \sigma ^{\uparrow \downarrow + \downarrow \uparrow}}
              {\sigma ^{\uparrow \uparrow + \downarrow \downarrow} +
               \sigma ^{\uparrow \downarrow + \downarrow \uparrow}} \:,
\end{equation}
for both beams fully polarized along the unit vector $\hat{n}$ normal to
the scattering plane. The asymmetry $A_{SS}$ is defined analogously,
but for both beams fully polarized along the unit vector $\hat{s}$
in the scattering plane and normal to the beam. Elastic $pp$-scattering is described by five
independent helicity amplitudes: two helicity conserving ones
$\phi _1$ and $\phi _3$, two double helicity-flip ones $\phi _2$ and $\phi _4$, 
and one single helicity-flip amplitude $\phi _5$ (see Ref.~\cite{buttimore} for definitions).
Each amplitude consists of hadronic and electromagnetic parts;
$\phi_i = \phi^{em}_i+\phi^{had}_i$.
The double spin asymmetries and $\Delta\sigma_T = \sigma^{\uparrow\downarrow} - \sigma^{\uparrow\uparrow}$, which are the subject of this paper, relate to the helicity amplitudes as follows \cite{buttimore}
\begin{eqnarray}
\label{eq:1}
A_{NN}\cdot\frac{d\sigma}{dt} &=& \frac{4\pi}{s^2}
    \{2|\phi_5|^2+{\rm Re}(\phi^*_1\cdot\phi_2-
    \phi^*_3\cdot\phi_4)\} \, ,\\ 
A_{SS}\cdot\frac{d\sigma}{dt} &=& \frac{4\pi}{s^2}
    {\rm Re}(\phi^*_1\cdot\phi_2+\phi^*_3\cdot\phi_4) \, ,\nonumber\\
\Delta\sigma_T &=&
    \frac{8\pi}{s}{\rm Im}(\phi_2)_{t=0} \, .\nonumber
\end{eqnarray}

In the small $t$ region the parameters $A_{NN}$ and $A_{SS}$
are sensitive to the interference between hadronic spin-flip amplitudes
$\phi^{had}_2$ and $\phi^{had}_4$, and the electromagnetic non-flip
amplitude. This provides a sensitive tool to study the spin dependence
of diffractive scattering at asymptotic energies and to search of
the hypothetical Odderon exchange \cite{leader}. 
Because Pomeron and Odderon have opposite C-parities, it is expected in leading order, that if  Pomeron and Odderon have the same asymptotic behaviour,
up to logarithms, they are out of phase by approximately 
90$^{\circ}$ \cite{eden}.
Therefore, if they couple to spin, their interference with the electromagnetic
non-flip amplitude will result in different $t$ dependences of the double
spin asymmetries 
(for a more detailed discussion of the $A_{NN}$ case see \cite{leader}).

\section{The Experiment}

\begin{figure}
\includegraphics[width=100mm]{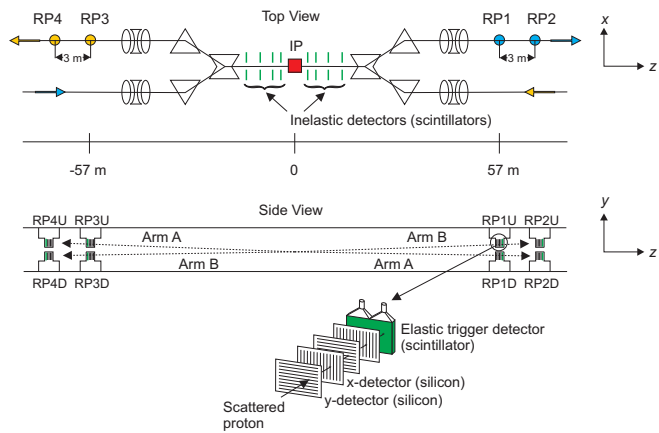}
\caption{\label{layout} Layout of the PP2PP experiment. Note the detector pairs RP1, RP2 and RP3, RP4 lie in different RHIC rings. Scattering is detected in either one of two arms: Arm A is formed from RP3U and RP1D.  Conversely, Arm B is formed from RP3D and RP1U. The coordinate system is also shown.}
\end{figure}

    The experiment pp2pp is located at the ``2 o'clock'' position of the
RHIC ring. The layout of the experiment is shown in Fig.~\ref{layout}.
The two protons collide at the interaction point (IP) and since the
scattering angles are small, the scattered protons stay within the beam
pipe until they reach the detectors. The measured coordinates are
related to the scattering angles by the beam transport matrix. The coordinates
are measured by silicon microstrip detectors (SSD) positioned just above and
below the beam orbits by insertion devices -- 
Roman Pots (RP) \cite{battiston}.
Each RP contains four planes of SSDs (two vertical and two horizontal)
to provide redundancy for the track reconstruction. High 
quality SSDs 
manufactured by Hamamatsu Photonics with sensitive area of 4.5 $\times $ 7.5~cm$^2$ and
pitch of 100~$\mu $m were used. 
The identification of elastic events is based on the collinearity criterion, hence it requires the simultaneous detection of the scattered
protons in a pair of RP detectors on either side of the IP.
More details on the experiment and the technique used can be found in \cite{PP2PPplb04,lynn}.

In addition, the inelastic event rates were monitored by eight scintillation
detectors, four on each side of the IP, positioned outside the beam pipe (cf. Fig. 1). To reduce background due to the beam halo the outer scintillators, labelled IP3, IP4 for the detectors on the side of the IP towards RP1,
and IN3, IN4 on the opposite side, were used for this analysis.
These detectors covered a pseudorapidity range of 4.2 $< \eta  <$ 5.3.
The inelastic trigger was defined as (IP3 OR IP4) AND (IN3 OR IN4) in coincidence
with the beam crossing signal from RHIC.

In order to prevent inelastic events from dominating the event rate
and causing excessive dead time, the inelastic triggers were prescaled 
by a factor of 500. The
data were recorded in an event-by-event mode, and broad offline cuts were made on 
the TDC and ADC to select events, yielding approximately 1400 inelastic events
for each of 50 bunch-pair crossings.

Each bunch in one RHIC ring (B) collides with a bunch in the other ring (Y)
to form different spin combinations. In this experiment, 13 different bunch-pairs
yielded collisions with B$^{\uparrow }$Y$^{\uparrow }$  bunches, 13 yielded B$^{\uparrow }$Y$^{\downarrow }$ bunches, 11 yielded B$^{\downarrow }$Y$^{\downarrow }$, and 13 yielded B$^{\downarrow }$Y$^{\uparrow }$
combinations

The inelastic event rate was used as one of the two methods to monitor
the relative luminosity for collisions of bunches with different polarizations. 
The product of intensities of colliding bunches was used as a second method
to monitor the relative luminosity. Both methods are described and compared in Chapter IV.

\section{Selection of elastic events}
	      
\begin{figure}
\includegraphics[width=100mm]{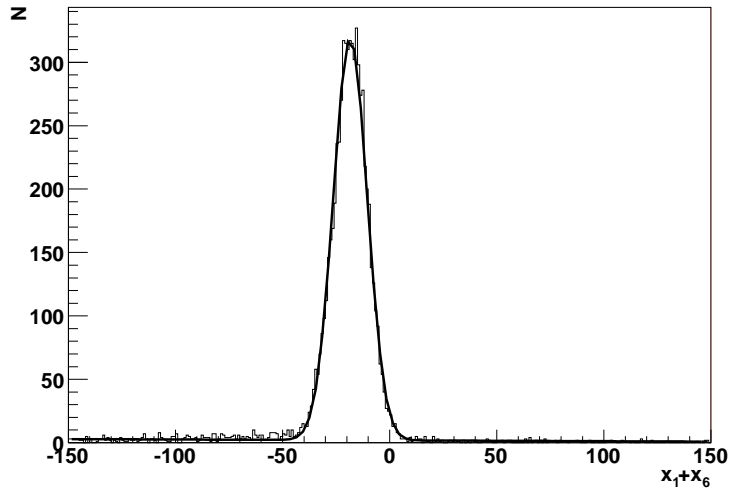}
\caption{Distribution of the sum of $x$-coordinates in a pair of silicon detectors on opposite sides of the IP. The coordinates are given in detector strip numbers.}
\label{fig2}
\end{figure}

Details on the identification of hits in the silicon detectors, hit clustering and selection of elastic events can be found in Ref.~\cite{PP2PPplb06}.
Because of the collinearity of the scattered protons a correlation between coordinates measured on each side of the IP is required. 
Hence, the main criterion to select the elastic scattering events was the hit coordinate correlation in the corresponding silicon detectors on the opposite sides of the  IP. 
An example of the distribution of the sum of x-coordinates measured
by a pair of detectors on both sides of the IP is shown in Fig.~\ref{fig2}.
One can observe that the background under the
elastic peak is small. To select an elastic event, a match of hit coordinates ($x$,$y$) from detectors on the opposite sides of the IP was required to be within $3\sigma$ for $x$- and $y$-coordinates.

The average detector efficiency was 0.98, and the upper bound of the elastic events loss  due to all criteria was $\le 3.5\%$. 
 
The background originates from particles of inelastic interactions, beam halo particles and products of beam-gas interactions. The estimated background fraction varies from 0.5\% to 9\% depending on the $y$-coordinate. Since in our analysis the coordinate area was essentially limited to $y > 30$ strips, the background in the final sample does not exceed 2\%. 
			    
\section{Determination of the double spin asymmetries}
			      
After the above cuts, a sample of 2.3 million elastic events was collected in the interval $0.010 \le -t \le 0.030$ (GeV/c)$^2$. 
The azimuthal angle dependence of the cross section for the elastic  collision of the vertically polarized protons is given by
\begin{equation}
2\pi\frac{d^2\sigma}{dt d\phi} = \frac{d\sigma}{dt} \cdot
(1 + (P_B + P_Y)A_N\cos{\phi} + P_B P_Y (A_{NN}\cos^2\phi + A_{SS}\sin^2\phi)) \, ,
\label{eq2}
\end{equation}
where $P_B$ and $P_Y$ are the beam polarizations, $A_{NN}$, $A_{SS}$ are the double spin asymmetries and $A_{N}$ is the single spin asymmetry.
The double spin raw asymmetry $\delta (\phi )$ is
\begin{eqnarray}
\nonumber
\delta (\phi) &=& P_B \: P_Y \: (A_{NN}\cos^2\phi+A_{SS}\sin^2\phi) \\
&=& \frac{N^{\uparrow\uparrow}(\phi)/L^{\uparrow\uparrow} + N^{\downarrow\downarrow}(\phi)/L^{\downarrow\downarrow} -
    N^{\uparrow\downarrow}(\phi)/L^{\uparrow\downarrow} - N^{\downarrow\uparrow}(\phi)/L^{\downarrow\uparrow}}
   {N^{\uparrow\uparrow}(\phi)/L^{\uparrow\uparrow} + N^{\downarrow\downarrow}(\phi)/L^{\downarrow\downarrow} +
    N^{\uparrow\downarrow}(\phi)/L^{\uparrow\downarrow} + N^{\downarrow\uparrow}(\phi)/L^{\downarrow\uparrow}} \, ,
\end{eqnarray}
where $L^{i,j}$ is the relative luminosity for the sum of bunches
with a given spin combination $\{i, j\} \in (\uparrow\uparrow, \uparrow\downarrow, \downarrow\uparrow, \downarrow\downarrow)$.
The raw asymmetry $\delta $ was calculated as a function of 
the azimuthal angle $\phi$ using $5^\circ$-bins in the three $t$-intervals 
$0.010 \le -t < 0.015$,  $0.015 \le -t < 0.020$, $0.020 \le -t \le 0.030$, as well as in the combined interval 
$0.010 \le -t \le 0.030$ (GeV/c)$^2$. 

To evaluate $\delta (\phi )$ two estimates of relative luminosities $L^{i,j}$
were used.
For the first one the rates of `inelastic' counters \cite{PP2PPplb04,lynn},
$N_{inel}^{i,j}$, summed
over bunch crossings with a given spin combination 
were used, $L_{inel}^{i,j} \sim \sum N_{inel}^{i,j}$.  As the second estimate the sums of
the machine
bunch intensities products for each
spin combination $L_{bint}^{i,j} \sim \sum I^i_B\cdot I^j_Y$ were used.
Because the relative statistical errors and relative errors on the luminosity
are comparable, they have to be combined for the determination of errors
of raw asymmetries.   
 
\begin{figure}
\includegraphics[width=100mm]{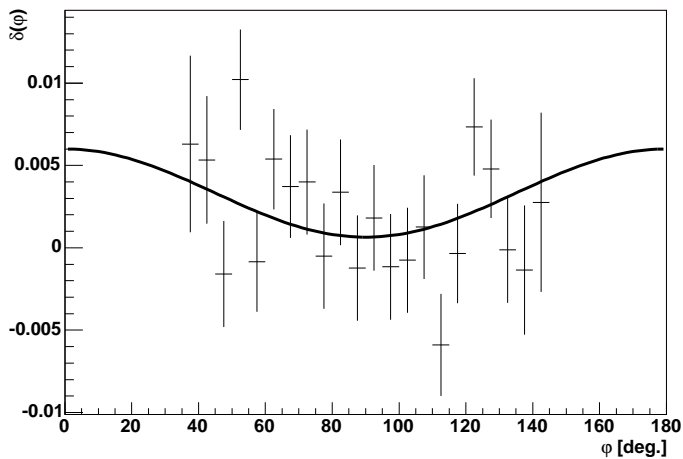}
\caption{The raw double spin asymmetry $\delta(\phi)$.}
\label{fig3}
\end{figure}

An example of the double spin raw asymmetry $\delta (\phi )$ distribution
for the whole $t-$interval 
is shown in 
Fig. \ref{fig3} together with the fitted function $P_B \: P_Y \: (A_{NN}\cos^2\phi+A_{SS}\sin^2\phi)$. $A_{NN}$ and $A_{SS}$ are the fit parameters and 
$P_B \cdot P_Y$ = 0.198$\pm$0.064. The displayed 
raw asymmetry was
obtained using bunch intensities for an estimate of the relative
luminosities.

In order to facilitate separation of contributions of the helicity
amplitudes $\phi_2$ and $\phi_4$ to the double spin asymmetries (cf. Eqs 2),
we performed also alternative fits to $\delta (\phi ) =P_B \: P_Y \: (a_1 +
a_2 \cos 2\phi )$ using $a_1 = (A_{NN}+A_{SS})/2$ and $a_2 = (A_{NN}-A_{SS})/2$ as fit parameters.

As a cross check the $\delta (\phi )$ distributions were obtained
using each of the two methods to estimate the relative luminosities and the results
of the fits to the $\delta (\phi )$ distributions were compared.
Both methods gave consistent results,
although  
with larger errors for normalization by the numbers of inelastic events.
Thus from now on we only present results obtained by using the bunch intensities. 

\section{Systematic Errors}

A detailed study of the errors on the relative luminosity estimated using
bunch intensities was carried out.
In the first stage of data processing the stability of the normalization
was checked by inspecting for each spin state $\{i, j\}$ the ratio
$N^{i,j}_{elast}/I_b^{i}\cdot I_Y^{j}$ as a function of the bunch crossing number.
The bunch crossings for which a significant deviation of the ratio 
from the average was observed were excluded from the analysis.
Next the $L_{bint}^{i,j} \sim I^{i,j} = \sum I^i_B\cdot I^j_Y$ summed over runs were evaluated for each position of detectors. The ratios between any pair of $I^{i,j}$'s are quite stable as a function of the position of detectors, at the level of $< 0.2 \%$.

For the method using bunch intensities to be applicable, it is
mandatory that $L_{bint}^{i,j} = k \,\cdot \, I^{i,j} = k \,\cdot \, \sum I^i_B\cdot I^j_Y$ and
the coefficient $k$ not to depend on the spin state.
The stability of $N^{i,j}_{elast}/I^{i,j}$ was analysed on a bunch-by-bunch
basis. The main result of this analysis is that the relative error
on the normalization is equal to $\sigma (k)/k = 0.0124\pm0.0015$.
From the analysis of the spread of $k$ on a run-by-run
basis we conclude that about half of the error on $k$ is due to the
accuracy of the intensity measurements. 

The systematic errors on the double spin asymmetries depend on the errors 
of the
positions of the beams in the $x$ and $y$ planes and on the transport matrix
elements, especially the effective lengths. The systematic errors due to these
two effects were estimated using the experimental sample and
assuming $\Delta x =$ 0.5 mm, $\Delta y =$ 1.0 mm, $\Delta L_{eff}^x/L_{eff}^x =$ 5\%
and $\Delta L_{eff}^y/L_{eff}^y =$ 5\% . 

The effect of these two sources on the systematic error
depends on a particular asymmetry; the uncertainty on the effective lengths
is the dominant contribution to
 the systematic error of $A_{NN}$ while the uncertainty on the
beam coordinates is dominating the one for $A_{SS}$.
The combined systematic errors for various double spin asymmetries are given
in Table I. 
The systematic errors are about half of the statistcal errors (including
the relative luminosity errors).
 
The polarization values of the proton beams were obtained from the 
accelerator polarimetry \cite{cad}. They were evaluated using $A_N$ measurements for elastic proton-carbon (pC) scattering at small $|t|$-values, in the range $0.01 - 0.02$ (GeV/c)$^2$. The details are described in Ref. \cite{carboncal}. During the period in 2003 when the present data were taken the beam polarizations were $P_Y=0.411\pm0.021$ and $P_B=0.481\pm0.027$. These values correspond to the present analysis, which 
includes bunches with all spin combinations.
The indicated errors are purely statistical.
In addition there is a systematic uncertainty of $13\%$,
due to the calibration of the pC polarimeter, which is correlated for both beams.  This gives for
the combined
statistical and systematic errors of the measurement of the product of the polarizations
$P_Y\cdot P_B = 0.198\pm0.064$. Thus, due to the uncertainties on the beam polarization
 the scale error on the measured asymmetries is equal to 32\% . 
 
\section{Results and Conclusions}

The results on the double spin asymmetries for the whole $t$-interval $0.010 \le -t \le 0.030$ (GeV/c)$^2$, at an average $<-t> = 0.0185$ (GeV/c)$^2$, are presented in Table I.
The most accurately determined asymmetry is $A_{SS} = 0.0035\pm0.0081$, which is consistent
with zero at 1$\sigma $ confidence level. The asymmetry $A_{NN} = 0.0298\pm0.0166$ as well as the combinations $(A_{NN}+A_{SS})/2 = 0.0167\pm0.0091$ and
$(A_{NN}-A_{SS})/2 = 0.0131\pm0.0096$ are also small and consistent with zero.

\begin{table}[h]
\caption{Double spin asymmetries $A_{NN}$, $A_{SS}$, $(A_{NN}+A_{SS})/2$ and
$(A_{NN}-A_{SS})/2$ for the $t$-interval $0.010 \le -t \le 0.030$ (GeV/c)$^2$ at $<-t> = 0.0185$ (GeV/c)$^2$.} \label{tab:DAresults}
\begin{center}
\begin{tabular}{|l|c|c|c|c|}
\hline
 &  $~~~~~~~A_{NN}~~~~~~~$ & $~~~~~~~A_{SS}~~~~~~~$ & $(A_{NN}+A_{SS})/2$ & $(A_{NN}-A_{SS})/2$ \\
\hline
$Asym$ & ~~0.0298 & ~~0.0035 & ~~0.0167 & ~~0.0131\\
\hline
$\Delta Asym$ (stat.+norm.) & $\pm$0.0166 & $\pm$0.0081 & $\pm$0.0091 & $\pm$0.0096 \\
\hline
$\Delta Asym$ (syst.) & $\pm$0.0045 & $\pm$0.0031 & $\pm$0.0034&$ \pm$0.0072 \\
\hline
$\Delta Asym$  due to $\Delta(P_Y \cdot P_B)$  & \multicolumn{4}{|c|}{$\pm$32.3 \%}\\
\hline
\end{tabular}
\end{center}
\end{table}

At collider energies one expects \cite{buttimore} the two helicity conserving amplitudes $\phi _1$ and $\phi _3$ to be equal, $\phi _1 \approx \phi _3 
\approx \phi _+ = (\phi _1 + \phi _3)/2$. In addition, if for simplicity
one assumes the Coulomb amplitude to be pure real, $\phi _+^{em} = -{\rm Im}\, \phi _+^{had}\cdot t_c/t$, where at our energy $t_c = -0.0013$ (GeV/c)$^2$, 
 the general formulae for the transverse double spin 
asymmetries (cf. Eq. 2) reduce to approximate ones:
\begin{eqnarray}
\label{eq:2}
A_{NN}\cdot {\rm Im}^2 \phi _+ & = & 2|\phi _5|^2 + {\rm Im}\, \phi _+\{{\rm Im}\, \phi _2 - {\rm Im}\, \phi _4 + (\rho - t_c/t)\cdot ({\rm Re}\, \phi _2 - {\rm Re}\, \phi _4)\} \, ,\nonumber\\
A_{SS}\cdot {\rm Im}^2 \phi _+  & = & {\rm Im}\, \phi _+\{{\rm Im}\, \phi _2 + {\rm Im}\, \phi _4 + (\rho - t_c/t)\cdot ({\rm Re}\, \phi _2 + {\rm Re}\, \phi _4)\} \, ,\nonumber\\
(A_{NN}+A_{SS})/2\cdot {\rm Im}^2 \phi _+ & = & |\phi _5|^2 + {\rm Im}\, \phi _+\{{\rm Im}\, \phi _2  + (\rho - t_c/t)\cdot {\rm Re}\, \phi _2 \} \, ,\nonumber\\
(A_{NN}-A_{SS})/2\cdot {\rm Im}^2 \phi _+ & = & |\phi _5|^2 - {\rm Im}\, \phi _+\{{\rm Im}\, \phi _4  + (\rho - t_c/t)\cdot {\rm Re}\, \phi _4 \} \, ,
\end{eqnarray}
where $\rho = {\rm Re}\, \phi _+ / {\rm Im}\, \phi _+$. For brevity,
in Eqs \ref{eq:2} and in the following we omitted the superscripts `had' for hadronic amplitudes.

As seen from the above formulae, taking $(A_{NN}+A_{SS})/2$ and $(A_{NN}-A_{SS})/2$
combinations allows one to select the amplitudes $\phi _2$ and $\phi _4$, respectively. 
Let us first assume that the phases of $\phi _2$ and $\phi _4$ are about the same
as the phase of the non-flip amplitude $\phi _+$.
Then the last term in the brackets can be neglected and the 1$\sigma $ 
upper limits
for the imaginary parts of $\phi _2$ and $\phi _4$ at 
$t = -0.185$ (GeV/c)$^2$ are:
${\rm Im}\, \phi _2/{\rm Im}\, \phi _+ < 0.028$ and 
$-{\rm Im}\, \phi _4/{\rm Im}\, \phi _+ < 0.026$, 
or in terms of the conventional
ratios $r_2 = \phi_ 2/(2\, {\rm Im}\, \phi _+)$ and $r_4 = -m^2\,\phi_ 4/(t\, {\rm Im}\, \phi _+)$ \cite{buttimore} they are ${\rm Im}\, r_2 < 0.014$ and ${\rm Im}\, r_4 < 1.25$.

A more precise limit on ${\rm Im}\, \phi _2$ at $t$ close to zero and therefore on $\Delta \sigma _T
= \sigma _{tot}^{\uparrow \downarrow} - \sigma _{tot}^{\uparrow \uparrow}$
can be obtained using the $t$-dependence of the asymmetry
$A_{SS}$ and extrapolating $A_{SS}$ to $t = -0.01$ (GeV/c)$^2$ where the term
containing the real parts of amplitudes vanishes.
For that purpose the corresponding experimental distributions $\delta (\phi )$ in the three $t$-intervals were
fitted with a function $P_B \: P_Y \: \{A_{SS}+(A_{NN}-A_{SS})\cos^2\phi\}$ 
with $A_{SS}$ as a fit parameter. Here the term $A_{NN}-A_{SS}$ was not fitted,
but calculated as a predefined function of $t$. At small $|t|$ this term is proportional to
$t$, $A_{NN}-A_{SS} = C\cdot t$, because of the 
kinematical factors in $\phi _5$ and $\phi _4$ resulting from
angular momentum conservation \cite{buttimore}.
The constant $C$ was calculated using the value of $A_{NN}-A_{SS}$
from the $\delta (\phi )$ fit for the combined $t$-interval (cf. Table I). 
The results of the fit of $A_{SS}$ in bins of $t$ are given in Table II.

\begin{table}[h]
\caption{Double spin asymmetry $A_{SS}$
in different intervals of $t$.} \label{tab:ASSresults}

\begin{center}
\begin{tabular}{|l|c|c|c|}
\hline
$-t$ interval (GeV/c)$^2$ &  0.010 - 0.015 & 0.015 - 0.020 & 0.020 - 0.030  \\
\hline
$-<t>$ (GeV/c)$^2$  interval &  ~~0.0127 & ~~0.0175 & ~~0.0236  \\
\hline
$A_{SS}$ & ~~0.0005 & ~~0.0076 & ~~0.0015\\
\hline
$\Delta A_{SS}$ (stat.)& $\pm$0.0071 & $\pm$0.0056 & $\pm$0.0061  \\
\hline
\end{tabular}
\end{center}
\end{table}

With the linear extrapolation to $t_0 = -0.01$ (GeV/c)$^2$ we obtain
 $A_{SS}(t_0) = 0.0037\pm0.0104$.
Neglecting the contribution of $\phi _4$ to $A_{SS}$ and the variation
of $\phi _2$ over the small range of $t$
one obtains ${\rm Im}\, \phi _2/{\rm Im}\, \phi _+ = 0.0037\pm0.0104$, 
${\rm Im}\, r_2 = 0.0019\pm0.0052$ and $\Delta \sigma _T = -0.19\pm0.53$ mb. The quoted errors are the quadratic sum of the statistical, normalization and
systematic errors and the error due to the beam polarizations uncertainty.

The existing experimental data on $\Delta \sigma _T$ at low energies are fairly well described by 
a Regge model with cuts \cite{andreeva}. Our result is consistent with the prediction of the model
for $\Delta \sigma _T$ at $\sqrt{s} = 200 \:\rm{GeV}$.

As seen from Eq. 5 the asymmetry $A_{SS}$ depends on the sum of the amplitudes 
$\phi _2$ and $\phi _4$ and this sum does not couple to
the leading and subleading Regge poles (Pomeron, Odderon, $\rho $, $\omega $, $f$, $a_2$) \cite{buttimore}. Thus one can expect that at $\sqrt{s} = 200 \:\rm{GeV}$
only Regge cuts may contribute. We consider the effect on $A_{SS}$ of a possible contribution of the Pomeron-Odderon
cut exchange in the $t$-channel as discussed in Refs~\cite{leader}
and \cite{rbrc}.
In case of such exchange the phase of the $\phi _2$ amplitude is shifted
by 90$^\circ $ relative to the amplitude $\phi _+$, and ${\rm Im}\, \phi _2 = -\rho \,{\rm Re}\, \phi _2$ and thus $A_{SS} \approx -t_c/t \cdot {\rm Re}\, \phi _2/{\rm Im}\, \phi _+$.
Using the $A_{SS}$ value at $t = -0.185$ (GeV/c)$^2$ (Table I) one obtains
${\rm Re}\, \phi _2/{\rm Im}\, \phi _+ = -0.050\pm0.130$ or ${\rm Re}\, r_2 = -0.025\pm0.065$.
Though this value is well consistent with zero it leaves wide room for a possible
Pomeron-Odderon cut contribution.

Theoretical predictions for double-spin asymmetries in elastic proton-proton
scattering at high energies and small momentum transfers 
have been recently presented in Ref.
\cite{rbrc}. The magnitudes of $A_{NN}$ and $A_{SS}$ have been estimated
using results from an earlier determination of the spin-couplings
of the leading Regge poles \cite{trspin04} and the required Regge cuts
were estimated using the absorptive Regge model.
As the Odderon spin coupling is totally unknown, the predictions
are given for various assumptions: (a) no Odderon, (b) weak Odderon spin
coupling - equal to that of the Pomeron, (c) strong Odderon spin coupling - equal
to the $\rho $ Reggeon spin coupling.
For none or a weak Odderon coupling 
the predicted values of the $A_{NN}$ and $A_{SS}$ asymmetries
are very small.
At $\sqrt{s} = 200 \:\rm{GeV}$ and $0.01 \leq |t| \leq 0.03$ $\GeVcSq$  
their values
are in the range 0.001 - 0.002.
On the contrary, for a strong Odderon spin coupling (like $\rho $) the 
double-spin asymmetries become significantly larger, at least by a factor of 10.
Our results on the $t$-dependence of $A_{SS}$ (cf. Table II) support 
predictions of Ref. \cite{rbrc} which 
assume none or a weak spin coupling of the Odderon.
 
In conclusion, these are the first measurements of the transverse double spin 
asymmetries and the first results on the double helicity-flip
amplitudes in the small $|t|$ region in elastic $pp$ scattering at collider energies. From the measured double spin asymmetries we determined the parameters ${\rm Im}\, r_2 = 0.0019\pm0.0053$ and $\Delta\sigma_T = -0.19\pm0.53$ mb, both being consistent with zero within errors.
We also estimated the upper limit on ${\rm Im}\,r_4$ which is ${\rm Im}\,r_4 < 1.25$.

Assuming the Pomeron-Odderon cut exchange one finds ${\rm Re}\, r_2 = -0.025\pm0.065$. The signs and central values of the real and imaginary parts of $r_2$ agree with expectations for Pomeron-Odderon cut exchange. Their magnitudes are consistent
with an assumption of about 5\% ratio of the cut amplitude to the dominant one.

\bibliography{apssamp} 
\begin{acknowledgments}
The research reported here has been performed in part under the US DOE
contract DE-AC02-98CH10886, and was supported by the US National
Science Foundation and the Polish Academy of Sciences. The authors
are grateful for the help of N.~Akchurin, D.~Alburger, P.~Draper,
R.~Fleysher, D.~Morse, Y.~Onel, A.~Penzo, and P.~Schiavon at various stages of the
experiment and for the support of the BNL Physics Department, Instrumentation Division, and the Collider-Accelerator Department at the RHIC-AGS facility. We would also like to thank T.~L.~Trueman for useful discussions.
\end{acknowledgments}

\end{document}